\documentclass{article}
\usepackage{emulateapj,apjfonts}
\input psfig

\newcommand{\etal}{{et~al.}}
\newcommand{\etnuk}{{et~al.}}

\newcommand{\Msun}{M_\odot}
\newcommand{\degr}{$^\circ$}
\newcommand{\mbh}{$M_\bullet$\ }
\newcommand{\mbhd}{$M_\bullet$}

\newcommand{\s}{$\sigma_e$\ }
\newcommand{\kms}{$\rm {km}~\rm s^{-1}$}

\newcommand{\mtli}{{\it M/L}$_{\rm I}$}
\newcommand{\vdm}{van~der~Marel}
\newcommand{\ts}{\thinspace}

\newcommand{\lapprox}{$_<\atop{^\sim}$}
\newcommand{\sd}{\newdimen\sa \sa=.1em  \ifmmode $\rlap{.}$''$\kern -\sa$
                                \else \rlap{.}$''$\kern -\sa\fi}
\newcommand{\m}{$\phantom{-}$}

\begin{document}

\lefthead{No BH in M{\ts}33}
\righthead{Gebhardt~\etal}

\title{M{\ts}33: A Galaxy with No Supermassive Black 
                 Hole\altaffilmark{1}}

\author{Karl Gebhardt\altaffilmark{2}, Tod R. Lauer\altaffilmark{3},
John Kormendy\altaffilmark{2}, Jason Pinkney\altaffilmark{4}, Gary A.
Bower\altaffilmark{3}, Richard Green\altaffilmark{3}, Theodore
Gull\altaffilmark{5}, J.~B.~Hutchings\altaffilmark{6},
M.~E.~Kaiser\altaffilmark{7}, Charles H. Nelson\altaffilmark{8},
Douglas Richstone\altaffilmark{4}, Donna Weistrop\altaffilmark{8}}

\altaffiltext{1}{Based on observations with the NASA/ESA {\it Hubble
Space Telescope}, obtained at the Space Telescope Science Institute,
which is operated by the Association of Universities for Research in
Astronomy, Inc.~(AURA), under NASA contract NAS5-26555.}

\altaffiltext{2}{Department of Astronomy, University of Texas, Austin,
Texas 78712; gebhardt@astro.as.utexas.edu, kormendy@astro.as.utexas.edu}
 
\altaffiltext{3}{National Optical Astronomy Observatories, P. O. Box
26732, Tucson, AZ 85726; lauer@noao.edu, gbower@noao.edu,
green@noao.edu}

\altaffiltext{4}{Dept. of Astronomy, Dennison Bldg., Univ. of
Michigan, Ann Arbor 48109; jpinkney@astro.lsa.umich.edu,
dor@astro.lsa.umich.edu}

\altaffiltext{5}{NASA/Goddard Space Flight Center, Code 681,
Greenbelt, MD 20771; gull@sea.gsfc.nasa.gov}

\altaffiltext{6}{Herzbreg Institute of Astrophysics NRC of Canada, V9E 2E7
Canada; john.hutchings@hia.nrc.ca}

\altaffiltext{7}{Department of Physics \& Astronomy, Johns Hopkins
University, Homewood Campus, Baltimore, MD 21218; kaiser@pha.jhu.edu}

\altaffiltext{8}{Department of Physics, University of Nevada, 4505
S.~Maryland Parkway, Las Vegas, NV 89154; cnelson@physics.unlv.edu,
weistrop@physics.unlv.edu}

\begin{abstract}

\pretolerance=3000 \tolerance=5000

Galaxies that contain bulges appear to contain central black holes
whose masses correlate with the velocity dispersion of the bulge.  We
show that no corresponding relationship applies in the pure disk
galaxy M{\ts}33.  Three-integral dynamical models fit {\it Hubble
Space Telescope} WFPC2 photometry and STIS spectroscopy best if the
central black hole mass is zero.  The upper limit is 1500 $\Msun$.
This is significantly below the mass expected from the velocity
dispersion of the nucleus and far below any mass predicted from the
disk kinematics. Our results suggest that supermassive black holes are
associated only with galaxy bulges and not with their disks.

\end{abstract}
 

\section{Introduction}

\pretolerance=3000 \tolerance=5000

The mass of a central black hole (BH) and the stellar orbital
structure in a galaxy are direct probes into the formation and
evolution history of both the galaxy and the BH.  The tight
correlation between BH mass \mbh and bulge velocity dispersion \s
(Gebhardt~\etal\ 2000a; Ferrarese \& Merritt 2000) establishes a
strong link between BHs and their hosts.  Many theories (e.{\ts}g.,
Silk \& Rees 1998; Haehnelt \& Kauffmann 2000; Ostriker 2000; Adams,
Graff, \& Richstone 2001) predict such a correlation.  As data and
theory improve, we hope to be able to discriminate between the
alternatives.

The \mbh -- \s correlation is derived using galaxies whose
bulge-to-total luminosity ratios $B/T$ range from \hbox{$\sim$\ts0.15
to 1.}  They even include galaxies with ``pseudobulges,'' i.{\ts}e.,
bulge-like, high-density central components that are believed to have
grown by the secular evolution of disks (see Kormendy 1993 for a
review).  Bulges and pseudobulges appear to satisfy the same \mbh --
\s correlation (Kormendy \& Gebhardt 2001; Kormendy~\etal\ 2001).
However, the above references show that BHs do not correlate with the
total gravitational potential of the disk. It therefore appears that
the main requirement for a supermassive BH is that a galaxy contain
some kind of bulge, or at least a mass gradient similar to one.
Still, pure disk galaxies have not been studied in nearly the same
detail as galaxies that contain bulges.

Observing pure disk galaxies is difficult, because they generally do
not have bright centers and because the small expected BH masses
require good spectral resolution.  The presence of nuclei -- compact
star clusters that are distinct from the disk -- have made it possible
to study a few bulgeless galaxies.  For example, based on its nuclear
velocity dispersion, Filippenko \& Ho (2001, see also Moran \etal\
1999) conclude that the dwarf Seyfert galaxy NGC 4395 has \mbh
\lapprox \ts80,000 $\Msun$.  However, a BH of only 100 $\Msun$ is
needed if the nucleus is radiating at the Eddington luminosity.
Therefore NGC 4395 does not put a significant strain on the \mbh -- \s
correlation.

M{\ts}33 is the best bulgeless galaxy to use to investigate the
relationship between BHs and disks.  It is a normal, large galaxy
(absolute magnitude $M_V = -19.0$), but it is only 0.8 Mpc away (van
den Bergh 1991).  The spatial scale is therefore unusually good,
0\sd26 pc$^{-1}$.  M{\ts}33 does not contain a bulge, but it does have
a bright nucleus.  Kormendy \& McClure (1993) measured its velocity
dispersion with the Canada-France-Hawaii Telescope and found that
$\sigma = 21 \pm 3$ km s$^{-1}$.  This implies an upper limit on the
mass of any central black hole of $M_\bullet$ \lapprox \ts$5\times
10^4$ $\Msun$.  M{\ts}33 was the first large galaxy in which a dead
quasar could strongly be ruled out.  Recently, Lauer~\etal\ (1998)
obtained WFPC2 photometry and suggested that $M_\bullet$ \lapprox
\ts$2\times 10^4$ $\Msun$, by reducing the upper limits for the core
radius of the nucleus. In this paper, we present {\it Hubble Space
Telescope\/} ({\it HST\/}) STIS spectroscopy of the nucleus obtained
with the 0\sd1 slit.  The spatial resolution is an order of magnitude
better than that in Kormendy \& McClure (1993).  This leads to a much
stronger limit on \mbhd.  Our best three-integral dynamical model has
\hbox{$M_\bullet$ = 0.}  Even our upper limit of 1500 $\Msun$ is
significantly below the value expected from the \mbh -- \s
correlation.


\vskip -8pt

\begin{figure*}[t]
\centerline{\psfig{file=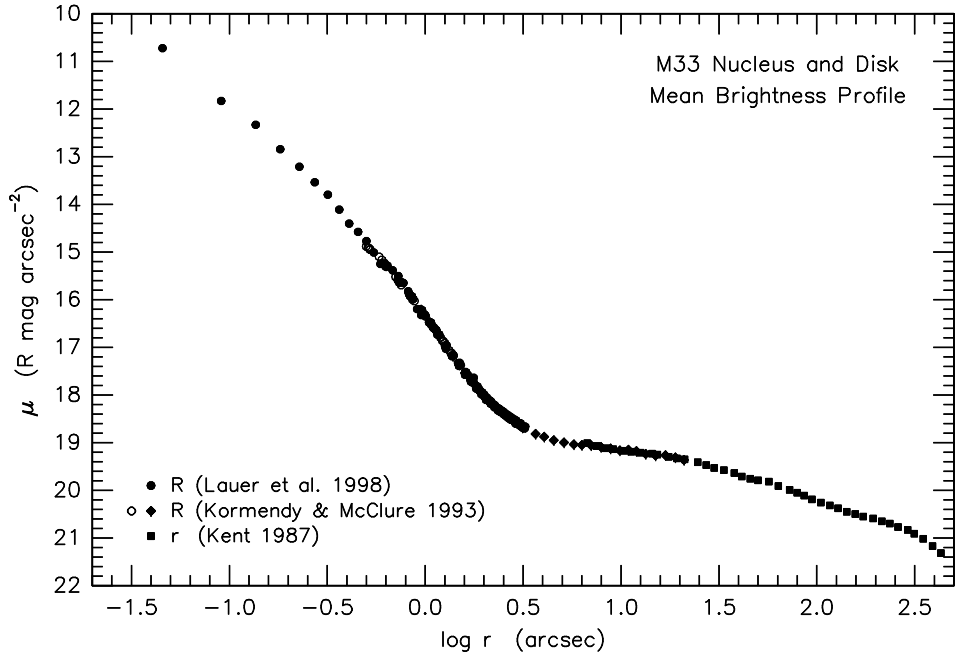,width=14cm,angle=-90}}
\vskip -55pt
\figcaption[gebhardt.fig1.ps]{Mean surface brightness profile of M{\ts}33
from Kormendy \& McClure (1993) but with ground-based data inside
0\sd5 replaced with {\it HST\/} photometry from Lauer \etal\ (1998).  
Their $V$- and $I$-band {\it HST\/} profiles have been used to 
synthesize the $R$-band profile plotted here.
\label{fig1}}
\end{figure*}
\vskip 0.3cm


It is important to emphasize how much M{\ts}33 differs from the
galaxies in which BHs have been found.  Figure 1 shows its brightness
profile and Figure 1 in Kormendy \& McClure (1993) shows an image of
the central 70$^{\prime\prime}$ \kern -.15pc $\times$ \kern -.15pc
113$^{\prime\prime}$.  There is no bulge; the exponential disk
dominates the structure in to~3$^{\prime\prime}$.  At the center,
there is a compact nuclear star cluster that is dynamically distinct
from the disk.  At absolute magnitude $M_B = -10.2$, this is like a
large globular cluster; it appears to be more closely related to
globular clusters than to bulges in the Fundamental Plane parameter
correlations (Kormendy \& McClure 1993).  The nucleus has a high
central stellar density, similar to that of M{\ts}31 and M{\ts}32
(Kormendy \& Richstone 1995; Lauer \etal\ 1998).  But its velocity
disperson is like that of a globular cluster, not like that of a
bulge.  Since mass $M \propto \sigma^2$, the difference in velocity
dispersion is the essential reason why we find BHs in M{\ts}31 and
M{\ts}32 but not in M{\ts}33.

\section{Photometry}

We use {\it HST} WFPC2 photometry from Lauer~\etal\ (1998), who
provide details of their reduction procedure.  The bandpasses are $V$
and $I$, and the spatial resolution of the final profiles is
0\sd025. Ground-based data at larger radii come from Kormendy \&
McClure (1993).  There is a color gradient in the nucleus; its center
is $\Delta\ts(V - I) = 0.15$ bluer than its outer parts (Fig.~2).  We
want to measure the mass profile by using the light profile to
represent the stellar potential, so we need to check how much the
color gradient affects the results.  It would be safest to use a
$K$-band profile, because this would be least sensitive to the
gradient in stellar population.  Since we don't have a $K$-band
profile at the desired spatial resolution, we construct one by using
the $V - I$ profile to estimate \hbox{$V - K$.}  Giant stars have a
monotonic and smooth relationship between $V - I$ and $V - K$ (Alves
2000, and references therein).  At $V - I \simeq 1$ (appropriate for
M{\ts}33), the relation is approximately linear, $\Delta$\ts$(V - K) =
3\ts\Delta$\ts$(V - I)$.  This relation provides the approximate
\hbox{$K$-band} profile shown in Fig.~2. To further check the effect
that color gradients have on the results, we have also run models
where we change the mass-to-light ratio as a function of radius
according to the $V-I$ profile; for these models we include a
variation of 20\% in the \mtli\ starting from the edge of the nucleus
decreasing linearly into the center. By either including a decrease in
the mass-to-light ratio towards the center or using a flatter surface
brightness profile (as suggested by the K-band profile), we are
increasing the likelihood of requiring a black hole. Our goal is to
place a robust and conservative upper limit on the black hole
mass. Dynamical models constructed using the measured $V$, measured
$I$, extrapolated $K$ profiles, and $I$ with varying \mtli\ profiles
all give very similar results. We conclude that the stellar population
gradient does not have a significant affect on our estimates of \mbhd.
We discuss the results from each of the profiles below.


\vskip 0.3cm

\psfig{file=gebhardt.fig2.ps,width=8.5cm,angle=0}
\figcaption[gebhardt.fig2.ps]{Surface brightness profiles of the
nucleus of M{\ts}33.  The $V$-band profile is the solid line, $I$-band
is the dashed line, and $K$-band is the dotted line.  We added 1.0 to
the $I$-band profile and 2.32 to the $K$-band profile to match them at
large radii.  The bluer-light profiles are steeper in the middle.
\label{fig2}}
\vskip 0.3cm



\begin{figure*}[t]
\centerline{\psfig{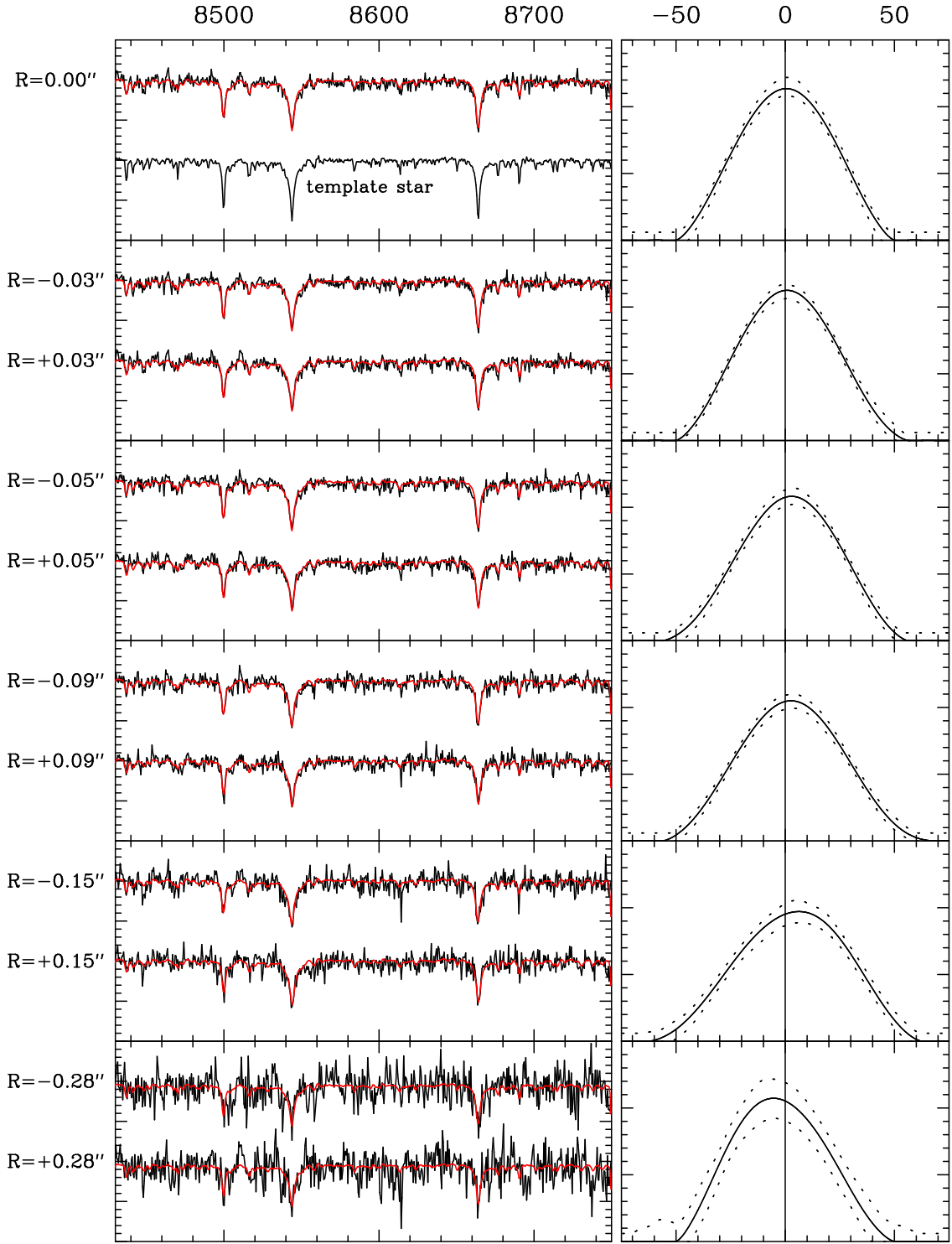}}
\figcaption[gebhardt.fig3.ps]{M{\ts}33 spectra and velocity profiles.
Each panel at left shows two radial bins.  The black line is
the galaxy spectrum used in the kinematic analysis.  The red line is the
spectrum of the template star convolved with the velocity profile.  The top
panel also shows the template star spectrum. The panels at right show the
velocity profiles and their 68\ts\% confidence bands.
\label{fig3}}
\end{figure*}


\begin{deluxetable}{ccccc}
\tablenum{1}
\tablewidth{27pc}
\tablecaption{Kinematics of the Nucleus of M{\ts}33}
\tablehead{
\colhead{Radius}   & 
\colhead{Velocity} &
\colhead{$\sigma$} &
\colhead{$h_3$}    &
\colhead{$h_4$}    \nl
\colhead{(arcsec)} & 
\colhead{(\kms)}   &
\colhead{(\kms)}   &
\colhead{}         &
\colhead{}         }
\startdata
 0.00 & $ 0.4\pm1.4$ &  $23.9\pm1.7$ &   $-0.00\pm0.03$ &  $-0.04\pm0.01$ \nl
 0.03 & $ 1.1\pm1.1$ &  $22.0\pm1.3$ & \m$ 0.02\pm0.03$ &  $-0.05\pm0.01$ \nl
 0.05 & $ 2.1\pm1.3$ &  $22.0\pm1.4$ &   $-0.04\pm0.03$ &  $-0.05\pm0.01$ \nl
 0.09 & $ 2.7\pm1.4$ &  $24.0\pm1.6$ & \m$ 0.00\pm0.02$ &  $-0.04\pm0.01$ \nl
 0.15 & $ 4.3\pm2.3$ &  $24.7\pm2.4$ &   $-0.02\pm0.03$ &  $-0.06\pm0.02$ \nl
 0.28 & $ 1.5\pm4.4$ &  $18.9\pm4.0$ & \m$ 0.05\pm0.04$ &  $-0.06\pm0.03$ \nl
 0.00 & $ 0.0\pm2.5$ &  $21.0\pm3.0$ &   \nodata        &  \nodata        \nl
 1.00 & $ 8.0\pm2.2$ &  $27.0\pm3.0$ &   \nodata        &  \nodata        \nl
\enddata
\tablecomments{The bottom two lines are ground-based measurements by Kormendy
               \& McClure (1993). They present only velocities and 
               dispersions; we use these with Gaussian LOSVDs in the modeling.}
\end{deluxetable}

\vfill

\section{HST STIS Kinematics}

\pretolerance=10000 \tolerance=10000

We measured the kinematics using the Space Telescope Imaging
Spectrograph (STIS; Woodgate~\etal\ 1998) centered at 8561 \AA.  Six
exposures at three different pointings had a total exposure time of
2.05 hours. The 0\sd1\ts$\times$\ts$52\arcsec$ slit and G750M grating
provided pixels of 0\sd05 $\times$ 0.55 \AA\ = 19 \kms.  The nucleus
of M{\ts}33 is slightly elongated (axial ratio $\simeq$ 0.85,
Lauer~\etal\ 1998); we put the slit along the major axis.
Contemporaneous flats were used to remove pixel-to-pixel sensitivity
variations and fringing.  We did not use the weekly dark frames
provided by STScI, because hot pixels come and go on timescales
shorter than a day.  Instead, we created a dark frame from the data
using the iterative procedure described in Pinkney~\etal\ (2001).  The
M{\ts}33 procedure differed slightly from normal: we have only 3
dither positions instead of 5, and the dither spacing is 4 pixels
instead of 20.  The steepness of the nuclear brightness profile
alleviates the light overlap problem caused by the small dither
spacing.  It takes 5 iterations to bring the residual galaxy light
level below 1\ts\% of the initial value, i.{\ts}e., acceptably below
the noise.  This process was used to create the dark frame that was
subtracted from the galaxy spectra.

The STIS spectra are well sampled in the spatial direction, so it was
possible to combine the dithered spectra without aliasing or the
introduction of any artifacts or smoothing that would degrade the
spatial resolution (Lauer 1999). We also choose to resample the
spectra in the spatial direction to provide the greatest flexibility
in sampling the kinematic parameters as a function of
radius. Combination of the spectra began with shifting the spectra to
a common spatial centroid to enable identification and elimination of
cosmic ray events. Cosmic ray events in one spectrum were replaced
with an average from the remaining spectra.  The ``cleaned'' spectra
were then shifted to a common centroid using sinc function
interpolation.  This allows well-sampled functions to be
reinterpolated without the introduction of smoothing.  The final
composite spectrum was sampled with 0\sd025 pixels. We also used a
cruder combination of the data that set the pixel scale at 0\sd05
pixels; in that case we find an upper limit on the black mass that is
a factor of two larger. Thus, the resampling procedure is critical as
it allowed superior use of the spatial information present in the
data.

The galaxy spectrum was radially binned according to the optimal
binning scheme used in the modeling.  That is, the observations and
models have the same radial bins.  We use nearly linear bins at small
radii and logarithmic bins at large radii (see Richstone~\etal\ 2001
for details).  The reduction procedure discussed in Pinkney~\etal\
(2001) was then applied to the extracted spectra.  Each spectrum was
divided by an estimate of its continuum.  This was based on binned
windows from which we measured a robust mean (the biweight) using the
highest 1/3 of the points.  We then linearly interpolated between the
mean values. The continuum division removes the need to flux calibrate
the spectra.

Obtaining the internal kinematic information requires a deconvolution
of the observed galaxy spectrum with a representative set of template
stellar spectra.  Both the deconvolution procedure and the choice of
templates stars affect the derived velocity profiles.  For the
template, we use the K giant star HR7615.  M{\ts}33 has younger stars
near its center than farther out in the nucleus, so using using only a
K III standard star may not be appropriate.  In fact, M{\ts}33's
nucleus clearly contains A stars (Gordon \etal\ 1999, and references
therein).  Fortunately, the Ca triplet region is not as sensitive to
template variations as is the more traditional Mg spectral region
(Dressler 1984). M{\ts}33 shows a small radial variation in Ca triplet
equivalent width from 10 \AA~at the center to 8~\AA~at 0\sd5.  Our
template star, HR7615, has a Ca triplet equivalent width of 8~\AA.
Garc\' ia-Vargas~\etal\ (1998) show that such a small difference in
equivalent width has a small effect on age and metalicity. In Section
5, we discuss the possibility of this equivalent width difference
affecting the deduced kinematics.

For the deconvolution, we reduce each spectrum using a
maximum-penalized-likelihood (MPL) technique that produces a
non-parametric line-of-sight velocity distribution (LOSVD).  First, we
guess initial velocity profiles in bins.  We convolve these profiles
with the template and calculate the residuals from the galaxy spectra.
The program then varies the velocity profile parameters -- i.{\ts}e.,
the bin heights -- to provide the best match to the galaxy
spectrum. The MPL technique is similar to that used in Saha \&
Williams (1994) and in Gebhardt~\etal\ (2000b). The complete set of
spectra and velocity profiles are shown in Figure 3.

Our models are axisymmetric, so at any absolute radius $|r|$, they
have the same LOSVDs on opposite sides of the center, except for the
sign of the velocity.  Thus, in order to maximize the signal-to-noise
ratios in our observations, we fit the same velocity profile shape to
both sides of the galaxy.  In Figure 3, we plot these velocity
profiles together with the two spectra at $\pm\ts|r|$.  We have also
measured the kinematics without symmetrizing.  As expected, we obtain
similar results but with larger uncertainties.  Since we are trying to
extract maximum information given low signal-to-noise levels in some
spectra, the symmetrized versions are preferred.

Monte~Carlo~simulations~are~used~to~measure~the~uncertainties in the
velocity profiles.  We convolve the template star with the measured
LOSVD to fit to the galaxy spectrum as illustrated by the red line in
Figure 3.  From this synthetic galaxy spectrum, we then generate 100
simulated observed spectra with noise, and we determine their velocity
profiles.  Each simulated spectrum contains random Gaussian noise with
the standard deviation given by the RMS of the initial fit.  The 100
simulated spectra and reductions provide a distribution of LOSVDs from
which we estimate the confidence bands. The dashed lines in the right
panels of Figure 3 represent the 68\ts\% confidence bands of the
distributions.  This bootstrap techniques estimates both the random
error and the bias of the estimator; however for these profiles the
bias is negligible. We do not test for the bias due to template
mismatch, but only estimator bias. The modeling described below uses
these velocity profiles directly; we do not use a parametric
representation of them. However, for convenience, Table~1 provides
parameters and uncertainties of Gauss-Hermite fits to the velocity
profiles.

Gebhardt~\etal\ (2000a) use the luminosity-weighted velocity
dispersion integrated out to the half-light radius of the bulge to
define the ``effective dispersion''.  Here, we follow the same
procedure for the nucleus. Integrating the STIS spectra from the
center to the edge of the nucleus, we find an effective velocity
dispersion of $\sigma_e = 24.0 \pm 1.2$ \kms.

\section{Three-Integral Models}

The 3-integral models are the same as in Gebhardt~\etal\ (2000b) and
are described in detail in Richstone~\etal\ (2001). The models are
orbit-based (Schwarzschild 1979) and constructed using maximum entropy
(Richstone \& Tremaine 1988). We run a representative set of orbits in
a specified potential and determine the non-negative weights of the
orbits to best fit the available data. The library of orbits is
convolved with the STIS PSF (as in Bower~\etal\ 2001) and binned
according to the STIS pixels. Maximizing the entropy helps to provide
a smooth phase-space distribution function.  Rix~\etal\ (1997) present
a similar code which has been applied to the spherical case;
Cretton~\etal\ (1999) and \vdm~\etal\ (1998) developed a fully-general
axisymmetric code that is also orbit-based.

Because of M{\ts}33's small dispersion and size we did not use the
standard setup parameters as in Pinkney~\etal\ (2001). Instead, we
designed the model bins to include the high-resolution sampling that
we used for the kinematic measurements. The central bins are 0\sd025
wide.

The model consists of 15 radial, 5 angular, and 13 velocity
bins. However, the force calculations use 4 times finer spatial
sampling in order to achieve energy conservation to better than 1\%
for most stars.  On average, each model contains about 3500 stellar
orbits. We tried a variety of inclinations for the nucleus from
edge-on to nearly face-on. In all cases we find a similar upper limit
for the black hole mass. Zaritsky, Elston, \& Hill (1989) measure an
inclination of 56\degr\ for the disk.

The data consist of six {\it HST} radial bins and two ground-based
bins with 13 velocity bins each.  However, the actual number of
independent measurements is less than 13 $\times$ 8, for several
reasons. First, the ground-based measurements add only six independent
values: a velocity and a dispersion at each of two positions. Second,
each STIS pixel is 19 \kms.  Given the 120 \kms\ region where the
velocity profile is non-zero, this implies about 6 independent
measurements. Third, the smoothing length in the LOSVD estimate
correlates some of the points. Thus, we have roughly 5 independent
measurements per the six HST velocity profiles combined with the 4
ground-based measurements, giving about 34 degrees of freedom.


\vskip 0.3cm

\psfig{file=gebhardt.fig4.ps,width=8.5cm,angle=0}
\figcaption[gebhardt.fig4.ps]{$\chi^2$ value as a function of black
hole mass using the $I$-band light for the stellar potential. The
points represent the models that we ran. The zero BH mass model is
shown as an arrow.
\label{fig4}}
\vskip 0.3cm


\section{Results}

For each of the three light profiles, we ran nine models with BH
masses ranging from zero to $1\times10^4\,\Msun$.  The total $\chi^2$
values are shown in Figure 4 as a function of BH mass using the
$I$-band light profile. For each BH mass, the best-fitting constant
stellar $M/L$ ratio is 0.70 in $I$ band. Figure 4 only shows the
one-dimensional $\chi^2$ values, and we do not plot these as a
function of $M/L$ since there is little information there. The minimum
$\chi^2$ value is around 27, which makes the reduced $\chi^2$ near unity.

Zero BH mass provides the best fit. Figure 4 only shows the results
for the $I$ band where we have allowed the \mtli\ to vary by 20\%, but
using any of the other light profiles also results in a minimum at
zero BH mass. The difference between the various light profiles is in
the upper limit for the black hole mass. For the $I$ band profile with
varying \mtli, $\Delta\chi^2 = 1$ at 1400 $\Msun$; for $I$ with
constant \mtli, the upper limit is 600 $\Msun$; using the $K$ profile,
the upper limit occurs at 1200 $\Msun$. We generally use a
$\Delta\chi^2$ equal to one to signify the extent of the viable
models, but, in this case, given the uncertainty of the light profile
we choose a conservative upper limit of 1500 $\Msun$.


\vskip 0.3cm

\psfig{file=gebhardt.fig5.ps,width=8.0cm,angle=0}
\figcaption[gebhardt.fig5.ps]{The first four Gauss-Hermite moments of
the velocity profiles for both model and data. The points with error
bars represent the data, with large circles the {\it HST} points and
small circles the two ground-based measurements of mean and
dispersion. There are two lines which represent models with black hole
masses of zero (red) and 15000 (black) $\Msun$. The point at the
smallest radii is actually centered at radius equal to zero.
\label{fig5}}
\vskip 0.3cm



\begin{figure*}[t]
\centerline{\psfig{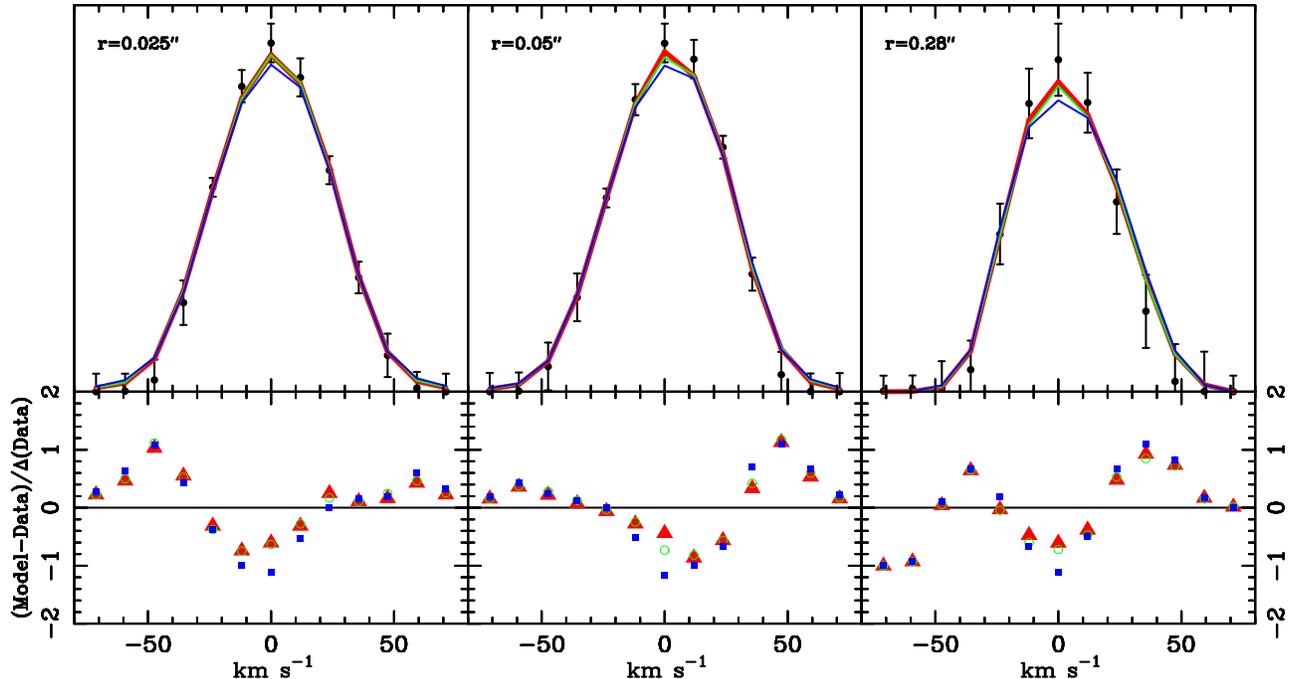}}
\figcaption[gebhardt.fig6.ps]{The comparison of the LOSVD from three
of the six STIS bins with model bins at radii as indicated. In the
upper panels, the solid circles with error bars represent the data
LOSVDs. The lines in the upper panel represent three different models,
and the lower panel shows their normalized residuals from the
data. The red line and triangles are the values from the zero BH
model, the green line and open circles are the model with a 2000
$\Msun$ BH, and the blue line and squares have a 7000 $\Msun$ BH.
\label{fig6}}
\end{figure*}


Figure 5 plots the first four Gauss-Hermite moments ($h_3$ and $h_4$
are measures of skewness and tail weight) of the velocity
distributions for both model and data. There are two different BH
models plotted in Figure 5, yet one would be very hard pressed to
discern any difference between them. In fact, the difference in
$\chi^2$ between the two models shown in Fig.~5 is 20, which
significantly excludes the large black hole mass model. However, we do
not use these Gauss-Hermite values during the fitting since the full
LOSVD determines the $\chi^2$ minimization. Thus, it is not optimal to
use these moments as a figure of merit to judge the quality of the
fit. Instead, one should inspect the comparison with the velocity
profiles. Figure~6 plots the LOSVD for three STIS bins with the same
three BH models. The difference between the various models can now be
seen. Furthermore, inspection of the other STIS velocity profiles
reveal similar results---i.e., the model with no BH consistently
provides the best fit. Only by looking at the total $\chi^2$
(Figure~4) can one judge the model fits.

The 3-Integral models also provide the stellar orbital distribution.
In order to constrain adequately the stellar orbits, we must have
two-dimensional spatial kinematic information. Major axis data allow
for too large a parameter space for the models. Based on comparison
with analytic models, the maximum entropy solution using only major
kinematic axis data does not recover the input orbital distribution
(Richstone~\etal\ 2001). Having only one additional position angle
helps considerably. Unfortunately, we only have major axis data for
M{\ts}33 and so do not report the orbital structure. However, with
only the major axis data, the estimate for the BH mass is unbiased.

\subsection{Possible Problems}

These 3-Integral models are limited by various assumptions. First, we
assume that the nucleus is spheroidal (elliptical isophotes with
ellipticity constant with radius). Second, the models are
axisymmetric, so we do not consider triaxial shapes. Third, we assume
specific radial forms for the $M/L$ variation.  Any of these
assumptions will reflect themselves as a bias in the models and
restrict the measured uncertainties. The most important is likely the
assumption of the $M/L$ variation. However, the color gradient is
rather small ($\Delta{\ts}V - I = 0.15$) and when we include a $M/L$
gradient, the results do not change significantly. This is apparent
since we find similar results when we use the extrapolated $K$-band
light profile.

No model can reproduce the $h_4$ values seen in the central few bins. We
have thoroughly checked the negative $h_4$ values in these bins using
extensive simulations to confirm that they are not an artifact of either
the data analysis or velocity profile estimation. They appear to
be a real feature in the data that is not reproducible in the
models. The absolute difference between the model $h_4$ of $-0.02$ and the
data $h_4$ of $-0.04$ is small but significant given the uncertainty. This
difference may either be due to use of an improper template or to our
lack of knowledge of the light distribution interior to our spatial
resolution. Since M{\ts}33's light profile is bluer in the center, template
mismatch is likely since we are only using a K giant star. An A star
will likely have a lower equivalent width for the Ca triplet,
depending on the metalicity (Garc\' ia-Vargas~\etal\ 1998); thus, if
the center of M{\ts}33 consists of A stars, a fit using a higher equivalent
width may result in an LOSVD with truncated wings, such as is
seen. However, the absolute difference is so small (0.02) that it will
have a negligible effect on our results.

Gordon~\etal\ (1999) advocate a strong foreground of absorbing dust
(V-band optical depth of 2), and we are not using an observed K-band
light profile. Instead, we have estimated the K light using the V and
I profiles. If the dust is distributed with a steep radial profile,
then our K-band estimation will be in error. However, the effect is to
make the upper black hole mass limit even tighter. As one includes
more dust near the center, the result is to steepen the actual K-band
profile and thereby increasing the amount of mass in the stellar
component. Thus, unless the dust is distributed in a clever way, the
upper limit that we derive on the black hole mass is robust.

We have assumed that the dynamical center of the galaxy is at the
nucleus. Photometrically, the center of the galaxy is difficult to
identify given the amount of dust; high spatial resolution IR imaging
is required to determine the photometric center. X-ray observations
(Colbert \& Mushotzky 1999) show a compact source within 2\arcsec\ of
the nucleus, which is well within the ROSAT HRI uncertainty; but given
that M33 does not appear to have a supermassive black hole, x-ray
observations are unlikely to point to the galaxy
center. Kinematically, the rotation curve is symmetric about the
nucleus giving weight that we are looking at the galaxy
center. However, the nature of the nucleus is unclear; it is similar
to a large core-collapse globular cluster in both photometric and
kinematic properties. Given the shallow potential well in the center
of M33, such a dense system would be free to wander. Thus, even though
there is no evidence to suggest a problem, a two-dimensional kinematic
map of the nuclear region would be very useful to pinpoint the
dynamical center.

\section{Conclusions}

The nuclear effective velocity dispersion of M{\ts}33 is
$\sim$\ts24~\kms.  If the mass of a central BH were related to this
dispersion of a nucleus like the known BHs are related to the
dispersions of their bulges; i.{\ts}e., if $M_{\bullet} =
1.3\times10^8~(\sigma_e/200)^{3.65}$ (Gebhardt~\etal\ 2001), then we
would expect a BH of \hbox{$M_{\bullet} = 5.6\times10^4$ $\Msun$.}
The corresponding relationship from Ferrarese \& Merritt (2000)
implies a mass of 7000 $\Msun$. The upper limit derived from the {\it
HST\/} data is 1500 $\Msun$ and the best-fit value is zero. Thus,
M{\ts}33 appears to lie below the \mbh~--~\s correlation for bulges.
This suggests that pure disk galaxies---even ones with nuclei---are
different from bulges and ellipticals in their BH properties. Figure~7
plots M33's upper limit in the \mbh~--~\s correlation. The line is
given by the above relation and the 68\% confidence bands are
determined through a bootstrap procedure (Gebhardt~\etal\ 2001).


\vskip 0.3cm

\psfig{file=gebhardt.fig7.ps,width=8.0cm,angle=0}
\figcaption[gebhardt.fig7.ps]{The \mbh -- \s correlation for the
galaxies from Kormendy \& Gebhardt (2001) including M33. The line is
the best fit to the BH detections; the relation is given by
$M_{\bullet} =1.3\times10^8~M_\odot\left(\sigma_e/200~{\rm
km~s^{-1}}\right)^{3.65}$. We have assumed for this plot that the
bulge of M33 is its nucleus.
\label{fig7}}
\vskip 0.3cm


There is a caveat.  Because there are no firm dynamical BH detections
in the mass range $10^3$ to $10^6$ $\Msun$, there is a big gap in the
\mbh~--~\s correlation between the smallest known supermassive BHs and
the mass limit in M{\ts}33.  That is, we do not know whether the
correlation extends down to the small masses of interest here; this is
not a problem: our observation may be a sign that it does not.  But
given the large gap, we also do not know that the correlation---if it
exists---is linear between $10^6$ and $10^3$ $\Msun$.  If the
correlation exists but is not linear, then our limit of \mbh \lapprox
\ts1500 $\Msun$ may not be below the correlation. Furthermore, the
confidence band in Figure~7 assumes a linear relation; thus,
deviations from a linear relation will have a significant affect on
the estimated uncertainties.

On the other hand, Kormendy \etal\ (2001) show that our $M_\bullet$
limit in M{\ts}33---and published masses and mass limits on BHs in
other disk galaxies---are far below the correlation of $M_\bullet$
with the total luminosities of their host galaxies.  M{\ts}33 is by
far the most extreme case.  There is a corresponding argument for the
\mbh~--~\s correlation.  The H{\ts}I rotation curve of the outer disk
of M{\ts}33 rises slowly from about 100 \kms\ at $r = 15^\prime$ = 3.5
kpc to about 135 \kms\ at $r = 77^\prime$ = 18 kpc (Corbelli \&
Salucci 2000).  If these are circular velocities in the gravitational
potential of an almost isothermal dark halo, then the corresponding
one-dimensional velocity dispersion is $\sigma \sim 85 \pm 10$ \kms.
If a BH in M{\ts}33 were related to the dark matter potential well
like known BHs are related to their bulge potential wells (both
represented by $\sigma_e$), then M{\ts}33 should contain a BH of mass
$M_\bullet \sim 6 \times 10^6$ $\Msun$.  Clearly it does not.  If
M{\ts}33 is typical of disk galaxies, then BH masses are not related
in the same way to every gravitational potential well (as measured by
$\sigma_e$).  In particular, they are not related to the potential
well of the disk and dark matter.  Taken together, our results and
those of Kormendy \etal\ (2001) strongly suggest that BH masses are
causally connected with the properties---luminosity, mass, and
especially velocity dispersion---of their host bulges and not with
their disks.

These results are still based on only a few, well-observed galaxies.
We therefore need to study the \mbh~--~\s correlation more fully at
the low-dispersion end.  This should provide further insight into the
fundamental differences between disk and bulge galaxies, and it may
help to tell us why such a correlation exists in the first place.

\acknowledgements 

K.G. was partially supported by NASA through Hubble Fellowship grant
HF-01090.01-97A awarded by the Space Telescope Science Institute.
This work was also supported by NASA Guaranteed Time Observer funding
to the STIS Science Team.

\end{document}